\documentclass[%
 reprint,
superscriptaddress,
 amsmath,amssymb,
 aps,
prl,
floatfix,
]{revtex4-2}

\usepackage{xspace}
\usepackage[normalem]{ulem}   

\usepackage{graphicx}
\usepackage{dcolumn}
\usepackage{bm}
\usepackage{hyperref}
\hypersetup{colorlinks=true, citecolor=blue, urlcolor=blue, linkcolor=blue}

\begin{document}

\preprint{APS/123-QED}

\title{Validation of the $^{10}$Be Ground-State Molecular Structure Using $^{10}\text{Be}(p,p\alpha)^{6}\text{He}$ Triple Differential Reaction Cross-Section Measurements}




\author{P.J. Li}
\email{lipengjie@impcas.ac.cn}
\affiliation{CAS Key Laboratory of High Precision Nuclear Spectroscopy, Institute of Modern Physics, Chinese Academy of Sciences, Lanzhou 730000, China}
\affiliation{Department of Physics, The University of Hong Kong, Hong Kong, China}

\author{D. Beaumel}
\email{didier.beaumel@ijclab.in2p3.fr}
\affiliation{Université Paris-Saclay, CNRS/IN2P3, IJCLab, 91405 Orsay, France}
\affiliation{RIKEN Nishina Center, Hirosawa 2-1, Wako, Saitama 351-0198, Japan}

\author{J. Lee}
\email{jleehc@hku.hk}
\affiliation{Department of Physics, The University of Hong Kong, Hong Kong, China}

\author{M. Assi\'e}
\affiliation{Université Paris-Saclay, CNRS/IN2P3, IJCLab, 91405 Orsay, France}

\author{S. Chen}
\affiliation{Department of Physics, The University of Hong Kong, Hong Kong, China}

\author{S. Franchoo}
\affiliation{Université Paris-Saclay, CNRS/IN2P3, IJCLab, 91405 Orsay, France}

\author{J. Gibelin}
\affiliation{LPC Caen UMR6534, IN2P3/CNRS, ENSICAEN, Université de Caen Normandie, 14050 Caen cedex, France}

\author{F. Hammache}
\affiliation{Université Paris-Saclay, CNRS/IN2P3, IJCLab, 91405 Orsay, France}

\author{T. Harada}
\affiliation{RIKEN Nishina Center, Hirosawa 2-1, Wako, Saitama 351-0198, Japan}

\author{Y. Kanada-En’yo}
\affiliation{Department of Physics, Kyoto University, Kyoto 606-8502, Japan}

\author{Y. Kubota}
\affiliation{RIKEN Nishina Center, Hirosawa 2-1, Wako, Saitama 351-0198, Japan}

\author{S. Leblond}
\affiliation{Department of Physics, The University of Hong Kong, Hong Kong, China}

\author{P.F. Liang}
\affiliation{Department of Physics, The University of Hong Kong, Hong Kong, China}

\author{T. Lokotko}
\affiliation{Department of Physics, The University of Hong Kong, Hong Kong, China}

\author{M. Lyu}
\affiliation{College of Physics, Nanjing University of Aeronautics and Astronautics, Nanjing 210016, China}
\affiliation{Key Laboratory of Aerospace Information Materials and Physics, Ministry of Industry and Information Technology, Nanjing 210016, China}

\author{F. M. Marqu\'es}
\affiliation{LPC Caen UMR6534, IN2P3/CNRS, ENSICAEN, Université de Caen Normandie, 14050 Caen cedex, France}

\author{Y. Matsuda}
\affiliation{Cyclotron and Radioisotope Center, Tohoku University, Sendai 980-8578, Japan}
\affiliation{Department of Physics, Konan University, Kobe 658-8501, Japan}

\author{K. Ogata}
\affiliation{Department of Physics, Kyushu University, Fukuoka 819-0395, Japan}
\affiliation{Research Center for Nuclear Physics (RCNP), Osaka University, Ibaraki 567-0047, Japan}

\author{H. Otsu}
\affiliation{RIKEN Nishina Center, Hirosawa 2-1, Wako, Saitama 351-0198, Japan}

\author{E. Rindel}
\affiliation{Université Paris-Saclay, CNRS/IN2P3, IJCLab, 91405 Orsay, France}

\author{L. Stuhl}
\affiliation{Center for Exotic Nuclear Studies, Institute for Basic Science, Daejeon 34126, Republic of Korea}
\affiliation{Center for Nuclear Study, University of Tokyo, Wako, Saitama 351-0198, Japan}

\author{D. Suzuki}
\affiliation{RIKEN Nishina Center, Hirosawa 2-1, Wako, Saitama 351-0198, Japan}

\author{Y. Togano}
\affiliation{Department of Physics, Rikkyo University, 3-34-1 Nishi-Ikebukuro, Toshima, Tokyo 172-8501, Japan}
\affiliation{RIKEN Nishina Center, Hirosawa 2-1, Wako, Saitama 351-0198, Japan}

\author{T. Tomai}
\affiliation{Department of Physics, Tokyo Institute of Technology, 2-12-1 O-Okayama, Meguro, Tokyo 152-8551, Japan}

\author{X.X. Xu}
\affiliation{CAS Key Laboratory of High Precision Nuclear Spectroscopy, Institute of Modern Physics, Chinese Academy of Sciences, Lanzhou 730000, China}
\affiliation{Department of Physics, The University of Hong Kong, Hong Kong, China}
\affiliation{Advanced Energy Science and Technology Guangdong Laboratory, Huizhou 516003, China}

\author{K. Yoshida}
\affiliation{Advanced Science Research Center, Japan Atomic Energy Agency, Tokai, Ibaraki 319-1195, Japan}

\author{J. Zenihiro}
\affiliation{Department of Physics, Kyoto University, Kyoto 606-8502, Japan}
\affiliation{RIKEN Nishina Center, Hirosawa 2-1, Wako, Saitama 351-0198, Japan}

\author{N.L. Achouri}
\affiliation{LPC Caen UMR6534, IN2P3/CNRS, ENSICAEN, Université de Caen Normandie, 14050 Caen cedex, France}

\author{T. Aumann}
\affiliation{Institut f\"ur Kernphysik, Technische Universit\"at Darmstadt, 64289 Darmstadt, Germany}
\affiliation{GSI Helmholtzzentrum f\"ur Schwerionenforschung, 64291 Darmstadt, Germany}

\author{H. Baba}
\affiliation{RIKEN Nishina Center, Hirosawa 2-1, Wako, Saitama 351-0198, Japan}

\author{G. Cardella}
\affiliation{Istituto Nazionale di Fisica Nucleare, Sezione di Catania, 95123 Catania, Italy}

\author{S. Ceruti}
\affiliation{INFN Sezione di Milano, 20133, Milano, Italy}

\author{A.I. Stefanescu}
\affiliation{Horia Hulubei National Institute for \text{R\&D} in Physics and Nuclear Engineering, IFIN-HH, 077125 \text{Bucureşti-Măgurele}, Romania}
\affiliation{Doctoral School of Physics, University of Bucharest, 077125 \text{Bucureşti-Măgurele}, Romania}
\affiliation{RIKEN Nishina Center, Hirosawa 2-1, Wako, Saitama 351-0198, Japan}

\author{A. Corsi}
\affiliation{IRFU, CEA, Universit\'e Paris-Saclay, F-91191 Gif-sur-Yvette, France}

\author{A. Frotscher}
\affiliation{Institut f\"ur Kernphysik, Technische Universit\"at Darmstadt, 64289 Darmstadt, Germany}

\author{J. Gao}
\affiliation{School of Physics and State Key Laboratory of Nuclear Physics and Technology, Peking University, Beijing 100871, China}

\author{A. Gillibert}
\affiliation{IRFU, CEA, Universit\'e Paris-Saclay, F-91191 Gif-sur-Yvette, France}

\author{K. Inaba}
\affiliation{Department of Physics, Kyoto University, Kitashirakawa-Oiwake, Sakyo, Kyoto 606-8502, Japan}

\author{T. Isobe}
\affiliation{RIKEN Nishina Center, Hirosawa 2-1, Wako, Saitama 351-0198, Japan}

\author{T. Kawabata}
\affiliation{Department of Physics, Osaka University, Toyonaka, Osaka 540-0043, Japan}

\author{N.Kitamura}
\affiliation{Center for Nuclear Study, University of Tokyo, Wako, Saitama 351-0198, Japan}

\author{T. Kobayashi}
\affiliation{Department of Physics, Tohoku University, Sendai 980-8578, Japan}

\author{Y. Kondo}
\affiliation{Department of Physics, Tokyo Institute of Technology, 2-12-1 O-Okayama, Meguro, Tokyo 152-8551, Japan}

\author{A. Kurihara}
\affiliation{Department of Physics, Tokyo Institute of Technology, 2-12-1 O-Okayama, Meguro, Tokyo 152-8551, Japan}

\author{H.N. Liu}
\affiliation{Key Laboratory of Beam Technology and Material Modification of Ministry of Education, College of Nuclear Science and Technology, Beijing Normal University, Beijing 100875, China}
\affiliation{IRFU, CEA, Universit\'e Paris-Saclay, F-91191 Gif-sur-Yvette, France}

\author{H. Miki}
\affiliation{Department of Physics, Tokyo Institute of Technology, 2-12-1 O-Okayama, Meguro, Tokyo 152-8551, Japan}

\author{T. Nakamura}
\affiliation{Department of Physics, Tokyo Institute of Technology, 2-12-1 O-Okayama, Meguro, Tokyo 152-8551, Japan}

\author{A. Obertelli}
\affiliation{Institut f\"ur Kernphysik, Technische Universit\"at Darmstadt, 64289 Darmstadt, Germany}

\author{N.A. Orr}
\affiliation{LPC Caen, ENSICAEN, Universit\'e de Caen, CNRS/IN2P3, F-14050 CAEN Cedex, France}

\author{V. Panin}
\affiliation{RIKEN Nishina Center, Hirosawa 2-1, Wako, Saitama 351-0198, Japan}

\author{M. Sasano}
\affiliation{RIKEN Nishina Center, Hirosawa 2-1, Wako, Saitama 351-0198, Japan}

\author{T. Shimada}
\affiliation{Department of Physics, Tokyo Institute of Technology, 2-12-1 O-Okayama, Meguro, Tokyo 152-8551, Japan}

\author{Y.L. Sun}
\affiliation{Institut f\"ur Kernphysik, Technische Universit\"at Darmstadt, 64289 Darmstadt, Germany}

\author{J. Tanaka}
\affiliation{RIKEN Nishina Center, Hirosawa 2-1, Wako, Saitama 351-0198, Japan}

\author{L. Trache} 
\affiliation{Horia Hulubei National Institute for \text{R\&D} in Physics and Nuclear Engineering, IFIN-HH, 077125 \text{Bucureşti-Măgurele}, Romania}

\author{D. Tudor}
\affiliation{Horia Hulubei National Institute for \text{R\&D} in Physics and Nuclear Engineering, IFIN-HH, 077125 \text{Bucureşti-Măgurele}, Romania}
\affiliation{Doctoral School of Physics, University of Bucharest, 077125 \text{Bucureşti-Măgurele}, Romania}

\author{T. Uesaka}
\affiliation{RIKEN Nishina Center, Hirosawa 2-1, Wako, Saitama 351-0198, Japan}

\author{H. Wang}
\affiliation{RIKEN Nishina Center, Hirosawa 2-1, Wako, Saitama 351-0198, Japan}

\author{H. Yamada}
\affiliation{Department of Physics, Tokyo Institute of Technology, 2-12-1 O-Okayama, Meguro, Tokyo 152-8551, Japan}

\author{Z.H. Yang}
\affiliation{School of Physics and State Key Laboratory of Nuclear Physics and Technology, Peking University, Beijing 100871, China}
\affiliation{RIKEN Nishina Center, Hirosawa 2-1, Wako, Saitama 351-0198, Japan}

\author{M. Yasuda}
\affiliation{Department of Physics, Tokyo Institute of Technology, 2-12-1 O-Okayama, Meguro, Tokyo 152-8551, Japan}

\date{\today}


\begin{abstract}

The cluster structure of the neutron-rich isotope $^{10}$Be has been probed via the $(p,p\alpha)$ reaction at 150 MeV/nucleon in inverse kinematics and in quasifree conditions. 
The populated states of $^{6}$He residues were investigated through missing mass spectroscopy. The triple differential cross-section for the ground-state transition was extracted for quasifree angle pairs ($\theta_{p}$, $\theta_{\alpha}$) and compared to distorted-wave impulse approximation reaction calculations performed in a microscopic framework using successively the Tohsaki-Horiuchi-Schuck-Röpke product wave-function and the wave-function deduced from Antisymmetrized Molecular Dynamics calculations. The remarkable agreement between calculated and measured cross-sections in both shape and magnitude validates the molecular structure description of the $^{10}$Be ground-state, configured as an $\alpha$-$\alpha$ core with two valence neutrons occupying $\pi$-type molecular orbitals.
\end{abstract}


\maketitle

\paragraph{Introduction}

The formation of structures inside a nucleus is an intriguing phenomenon driven in part by correlations coming from the details of the nucleon-nucleon interaction. Among the different partitioning possibilities within a given nucleus, $\alpha$-clustering has always been considered the most favorable due to the large binding energy of the $\alpha$-particle and its inert character. Consequently, nuclei composed of an integer number of $\alpha$-particles (the so-called self-conjugate nuclei) have initially focused clustering studies. The Ikeda diagram~\cite{ikeda_the_1968}, which was proposed at the end of the 1960s, conveys the idea that the cluster degrees of freedom appear in the vicinity of the alpha emission threshold. The famous Hoyle state~\cite{Hoyle1954zz}, the second 0$^{+}$ state of $^{12}$C which plays a key-role in the nucleosynthesis of elements heavier than helium is a typical example of such a cluster state~\cite{freer_clustered_2007}. It is located at an energy just above the $3\alpha$ threshold in $^{12}$C. Its basic structure in three $\alpha$-particles is established, but its detailed nature is still an object of study. For example, this state can be described as a condensate of $\alpha$-particles occupying a large volume using a wave function of Tohsaki-Horiuchi-Schuck-Roepke (THSR) type~\cite{tohsaki_alpha_2001}. Calculations in the antisymmetrized molecular dynamics (AMD) approach confirm the significant spatial extension of this state and suggest a structure dominated by a "loose" configuration of 3$\alpha$~\cite{freer_microscopic_2018}. In contrast, the ground state of $^{12}$C is described in the above models as rather compact, with mean-field like structure, in accordance with the idea of the Ikeda diagram.

As compared to self-conjugate nuclei, the situation is different in neutron-rich light nuclei. For example, strong indications exist that in low-lying states of Be and B isotopes, including the ground-state, adding neutrons to an $N=Z$ core leads to spatially extended molecular-like structures in which valence neutrons orbit around the core composed of $\alpha$-particles~\cite{kanada-enyo_structure_2001,vonoertzen_nuclear_2006}. A typical case is represented by the neutron-rich ($N>4$) Be isotopes which were initially described as systems of $2\alpha + \text{X}n$, the two alphas forming a dumbbell-shaped core in the intrinsic frame and X being the number of excess neutrons occupying molecular orbitals around this core~\cite{vonoertzen_nuclear_2006}. The ground state of $^{9}$Be, the only stable beryllium isotope, may be described as a 3-body $\alpha$-$n$-$\alpha$ molecular structure. The observed rotational band built on the $^{9}$Be ground-state is well understood in terms of a $\pi$ molecular orbital, while the band built on  the first ($1/2^{+}$) excited state at 1.68 MeV can be connected to a $\sigma$-type molecular structure~\cite{von_oertzen_two-center_1996,von_oertzen_dimers_1997,DESCOUVEMONT2002463}. This description of the cluster structure initially elaborated in molecular orbital models was later confirmed by mean-field type approaches, namely AMD, from which the cluster structures emerge without the existence of clusters being presupposed. 
The next neutron-rich Be isotope is $^{10}$Be, an unstable nucleus whose structure is also expected to exhibit a marked molecular character. Experimental studies of this nucleus have revealed four rotational bands, corresponding to various cluster structures for excited states~\cite{von_oertzen_two-center_1996,von_oertzen_dimers_1997,Soic_1996,Curtis_decay_2001,DESCOUVEMONT2002463,milin_sequential_2005,freer_alpha_2006}. However, little is known about the cluster structure of its ground state, apart from its associated  rotational band. 
Breakup and neutron-removal reactions~\cite{Ashwood_helium_2004} provided evidence for the existence of $^x$He+$^{A-x}$He di-cluster structures in $^{10,12,14}$Be. The charge radius, which can be measured precisely, is directly related to the density distribution of protons, though not probing directly the cluster structure of ground-states.
Among the beryllium isotopes, $^{10}$Be exhibits the smallest charge radius(2.36 fm)~\cite{ANGELI201369}, which is consistent with AMD calculations that predict a minimum for $N=6$~\cite{kanada-enyo_proton_2015}. In these calculations, the ground state of $^{10}$Be is described as a $2\alpha + 2n$ configuration, in which valence neutrons occupy the molecular attractive $\pi$ orbital which produces a more compact $2\alpha$ core, at variance with $\sigma$ orbitals~\cite{kanada-enyo_structure_2001}. It should be noted that this configuration is also predicted in Density Functional Theory (DFT) calculations~\cite{ebran_density_2014}, a more general framework within which a large number of properties of nuclei can be reproduced.  Namely, the dumbbell shape structure of the alphas as well as the ring-shape $\pi$-type orbit of the neutrons emerge from the mean-field. 

\begin{figure}[tpb]
        \includegraphics{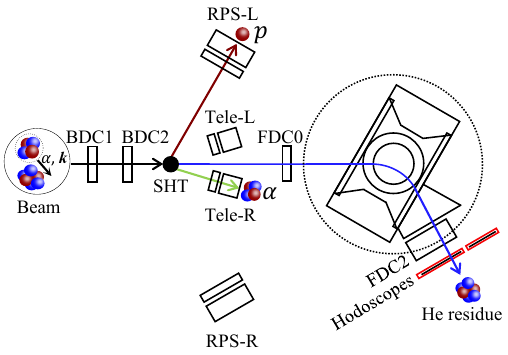}
        \caption{Schematic view of $^{10}$Be$(p,p\alpha)$$^{6}$He reaction setup.}
        \label{fig:fig-setup}
\end{figure}

In order to directly probe the spatial extension of $\alpha$-clusters in the ground state of $^{10}$Be and thus establish its overall $\alpha$-cluster molecular structure, we implemented a method based on cluster knockout reactions of the ($p,p\alpha$) type. These reactions have been studied extensively with proton beams on stable targets from the 70's until today, and were used essentially to extract the $\alpha$-cluster spectroscopic factors. These spectroscopic factors do not provide direct information on the spatial distribution of $\alpha$-particles in the ground-states of nuclei. 
On the other hand, the recent theoretical work ~\cite{lyu2018} has demonstrated the high sensitivity of the triple differential cross-section (TDX) of the $^{10}$Be($p,p\alpha$) reaction to the $\alpha$-particle wave function hence to the geometric configuration of the $\alpha$-clusters in the ground-state of $^{10}$Be. 
In the present work we have performed for the first time the measurement of the TDX of the ($p,p\alpha$) reaction in inverse kinematics using a beam of unstable nuclei of $^{10}$Be and in kinematical conditions covering the recoilless condition~(zero momentum transfer). The measured TDX was further compared to reaction calculations carried out in a microscopic framework, including in particular the microscopic cluster wave function so as to infer the $\alpha$-cluster molecular structure of the $^{10}$Be ground-state.
\paragraph{Experiment}
The experiment was performed at the Radioactive Isotope Beam Factory (RIBF) at RIKEN.
A secondary beam of $^{10}$Be was produced at an energy of approximately 150 MeV/nucleon through projectile fragmentation of a 230 MeV/nucleon $^{18}$O beam impinging on a 15-mm-thick Be target
and purified using the BigRIPS fragment separator~\cite{Kubo2003}. The average $^{18}$O beam intensity was 500 pnA
and the produced $^{10}$Be beam intensity was of ~5$\times$10$^{5}$ particles/second with a purity higher than 90$\%$. $^{10}$Be beam particles were identified on an event-by-event basis. Figure~\ref{fig:fig-setup} schematically shows the main components of the experimental setup around the secondary target. Beam ions were tracked by a set of two multiwire drift chambers~(MWDC), BDC1 and BDC2 placed upstream of the target chamber. To minimize multiple scattering of recoil protons from the $^{10}$Be($p,p\alpha$) reaction in inverse kinematics, a 2-mm-thick pure solid hydrogen target~(SHT)~\cite{Matsuda2011} was used as the reaction target.  
Recoil protons were detected using the Recoil Proton Spectrometer (RPS) described in~\cite{Chebotaryov_2018_proton,matsuda_elastic_2013} in a two-arm configuration set at 60$^{\circ}$ with respect to the beam axis. Each arm was composed of three stages (MWDC, plastic scintillator and NaI(Tl) rods) providing
position and energy measurement which were used to reconstruct the scattering angle and total energy of the recoil protons. Given the energy range of the recoil protons for the reaction of interest in the present measurement (25-100 MeV), data from the elastic and inelastic channels which produce protons in the relevant energy range were used to perform the energy calibration of RPS. 
Knocked-out $\alpha$-clusters were measured by two telescopes composed of double-sided strip silicon detector (DSSD) of 62$\times$62 mm$^2$ active surface backed by CsI(Tl) modules from the FARCOS array~\cite{verde_farcos_2013} placed in the horizontal plane to cover the angular range 4$^{\circ}$-12$^{\circ}$. The scattering angle of Helium residues emitted along the beam direction was determined using the MWDC (referred as FDC0 in Figure~\ref{fig:fig-setup}) placed downstream the SHT, and their identification  was performed using the SAMURAI spectrometer and its standard plastic hodoscopes~\cite{Kobayashi2013}.

\paragraph{Results}

\begin{figure}[htpb]
		\centering
		\includegraphics[width=\linewidth]{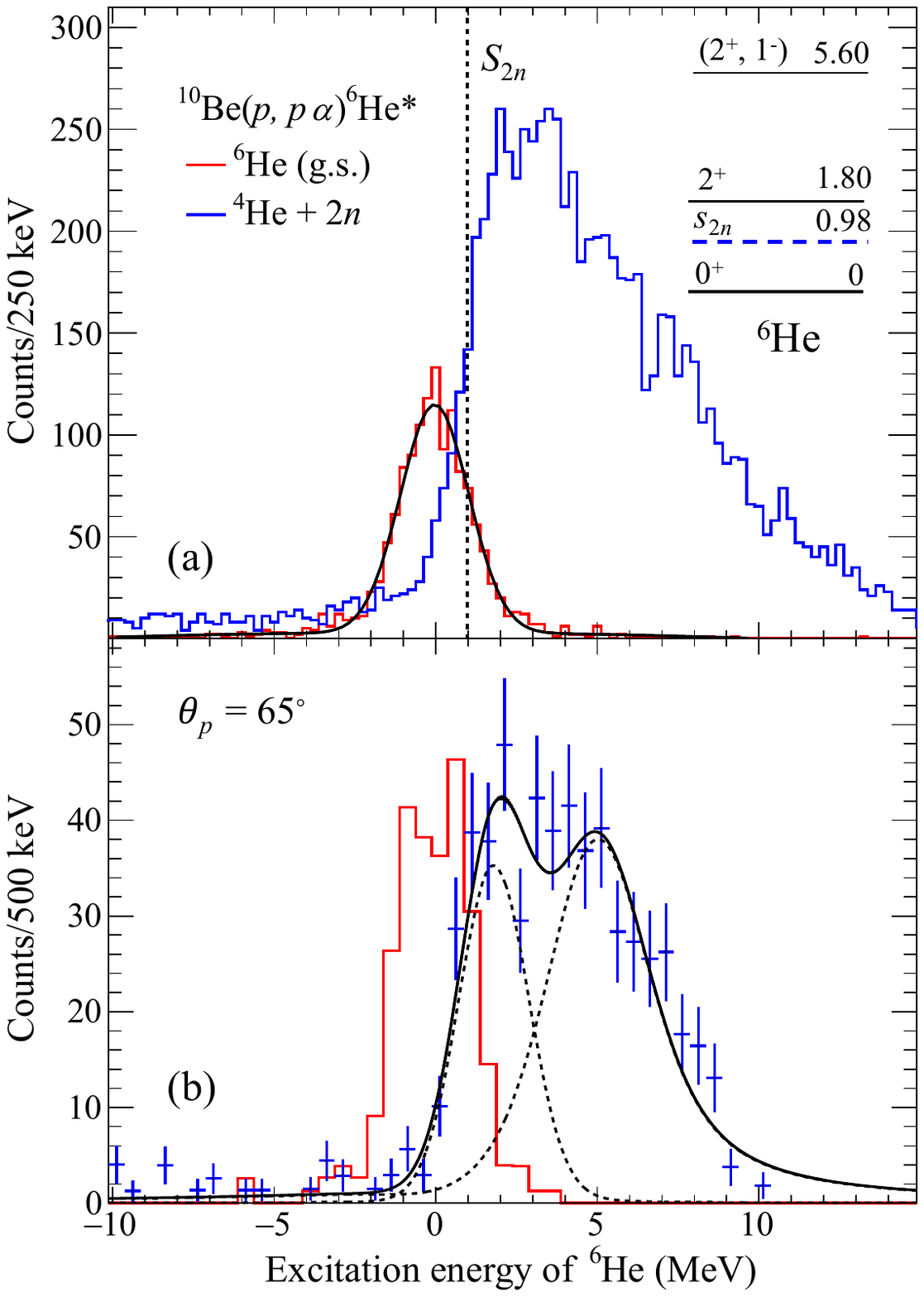}
		\caption{Excitation energy spectra for $^{10}$Be$(p,p\alpha)$$^{6}$He$^{*}$ reaction.
		(a) integrated over the full solid angle covered by proton and cluster detectors. The vertical dashed line indicates the 2-neutron separation energy(0.975 MeV).
		The red and blue solid lines show spectra corresponding to events gated by $^{6}$He and $^{4}$He residues, respectively. 
		(b) same for events corresponding to the quasifree angle pair ($\theta_{p}/\theta_{\alpha} = 65^{\circ}/-7.7^{\circ}$) with an angular bin size of $\pm 1 ^{\circ}$. See text for details.
		 }
		\label{fig:fig-Ex}
\end{figure}

Figure~\ref{fig:fig-Ex}(a) displays the excitation energy spectra in $^{6}$He for the $^{10}$Be$(p,p\alpha)$ reaction obtained from the measured energy and angle of both the recoil proton and $\alpha$-cluster. For the purpose of the energy calibration of the telescopes used for the detection of knocked-out clusters, secondary beams of $\alpha$-particles were produced using BigRIPS in a dedicated run for obtaining the energy calibration while checking the homogeneity of the CsI crystals response. 
The present triple coincidence measurement produces spectra with a rather low background. The ground state (g.s.) being the only bound state in $^6$He, the g.s.$\rightarrow$g.s. transition is easily separated by setting a gate on the $^{6}$He residues in SAMURAI. The corresponding peak (red histogram) 
is well fitted by a Gaussian function centered at (-0.02 $\pm$ 0.03) MeV with a missing mass resolution $\sigma$ = 1.06 MeV, which validates the calibrations of the proton and $\alpha$-clusters detectors. 
The spectrum gated by $^4$He residues in SAMURAI (blue histogram) corresponds to excited states populated in the $^{10}$Be$(p,p\alpha)$$^{6}$He$^{*}\rightarrow$$^{4}$He$+$2$n$ reaction channel. 
The resolution does not allow a clear identification of the populated states. 
The well-known 2$^{+}$ state can be seen at 1.797 MeV~\cite{tilley_energy_2002}, but not well separated from other low-lying resonant states in the 2-6 MeV region reported in Ref.~\cite{tilley_energy_2002,mougeot_new_2012,fortune_jensuremathpi_2014,mandaglio_first_2014,gurov_highly_2015,Chernyshev2018KnE}.
A less pronounced resonance at around 3.5 MeV is inferred to be the 2$_{2}^{+}$ state, observed in Ref.~\cite{gurov_highly_2015}, and predicted by theoretical calculations~\cite{ogawa_investigation_2020,myo_resonances_2007}.
Figure~\ref{fig:fig-Ex}(b) will be introduced in a later section.

\begin{figure}[htpb]
		\centering
		\includegraphics[width=\linewidth]{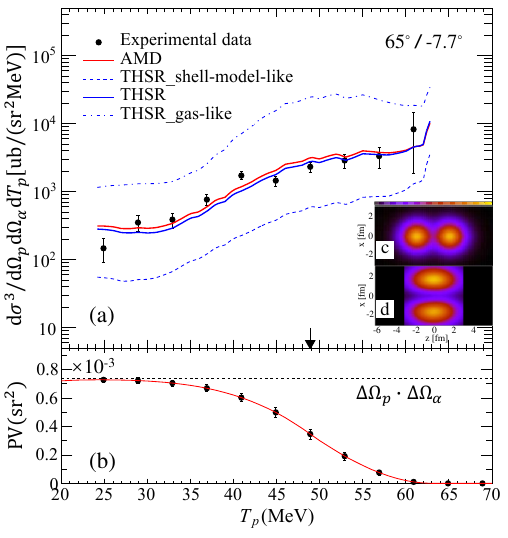}
		\caption{(a) TDX distribution  of $^{10}$Be$(p,p\alpha)$$^{6}$He(g.s.) reaction extracted at the coplanar angle pair $\theta_{p}/\theta_{\alpha} = 65^{\circ}/-7.7^{\circ}$ which is chosen to include recoilless condition of the residual nucleus. The arrow indicates the proton kinetic energy $T_{p}$ at the recoil-less condition. The red and blue solid lines are the DWIA predictions using the THSR and AMD structural models, respectively. The blue dashed-line (dot-dashed-line) is the TDX for an artificial state describing the compact shell-model-like(loosely bound gas-like) structure of $^{10}$Be nucleus~\cite{lyu2018}. (b) The corresponding phase volume distribution. The horizontal dashed line indicates the product of solid angles $\Delta \Omega_p\cdot \Delta \Omega_{\alpha}$. (c) and (d) are the density distribution of the protons and valence neutrons in the ground state of $^{10}$Be predicted by the THSR model, adapted from Ref.~\cite{lyu2018,lyu2016}. }
		\label{fig:fig-TDX}
\end{figure}

The experimental TDXs for the $(p,p\alpha)$ reaction were extracted for the coplanar angle pairs $(\theta_{p},\theta_{\alpha} )$  chosen to include the zero recoil momentum condition of the residual nucleus.
The experimental TDX is given in the unit of $\mu$b/(sr$^{2}\cdot$MeV) in the laboratory system.
For a given angle pair $\left(\theta_{p}, \theta_{\alpha}\right)$, the TDX for a given transition can be written as
\begin{equation}
\begin{aligned}
\frac{\text{d}^{3} \sigma^{exp}}{\text{d} T_{p} \text{d} \Omega_{p} \text{d} \Omega_{\alpha}}
& = \frac{\Delta N(T_{p})}{\varepsilon_{p\alpha}(\theta_{p})  N_{t}N_{b} \Delta T_{p}
\cdot PV(T_{p})
}, 
\end{aligned}
\label{eq:TDX}
\end{equation}
where index $p$ and $\alpha$ stand for the outgoing proton and $\alpha$-particle, respectively;
$T_{p}$ is the proton kinetic energy; $\Delta N$ is the number of counts in an energy bin $\Delta T_{p}$;
$\varepsilon_{p\alpha}\left(\theta_{p}\right)$ is the overall efficiency of detecting a $p$-$\alpha$ pair at $\theta_{p}$ and $\theta_\alpha$ obtained from the NPTool simulation~\cite{matta_nptool_2016};
$N_{t}$ is the number of protons per unit area of the SHT;
$N_{b}$ is the number of incident beam particles;
the phase volume term $\text{PV}(T_{p})$ corresponding to the portion of the  $\Delta \Omega_p\cdot \Delta \Omega_{\alpha}$ volume kinematically allowed can be defined in discrete form by :
\begin{align}
\!\text{PV}(T_{p})\!=\!\sum_{\theta_{p}}\!\sum_{\varphi_{p}}\!\sum_{\theta_{\alpha}}\!\sum_{\varphi_{\alpha}}\!\sin\!\theta_{p}\Delta\theta_{p}\Delta\varphi_{p}\!\sin\!\theta_{\alpha}\Delta\theta_{\alpha}\Delta\varphi_{\alpha},
\end{align}
where the summation ranges over $\theta_{p}$, $\theta_{\alpha}$, $\varphi_{p}$, and $\varphi_{\alpha}$ are restricted to satisfy the energy-momentum conservation.

Figure~\ref{fig:fig-TDX}(a) shows the extracted experimental TDXs as a function of the recoil proton kinetic energy $T_{p}$ for the  $^{10}$Be$(p,p\alpha)^{6}$He(g.s.) reaction at the angle pair $(\theta_{p}/\theta_{\alpha} = 65^{\circ}/-7.7^{\circ}$) compared with the reaction calculations discussed below. 
The arrow in the plot indicates the value of $T_{p}^{RL}$ corresponding to the recoilless condition.
The error bars correspond to the square-root of the quadratic sum of statistical uncertainties and those on the PV induced by the error on the scattering angles.
We note that the shape of the TDX distribution in inverse kinematics is heavily influenced by the PV term shown in Fig.~\ref{fig:fig-TDX}(b), which appears in the denominator of Eq. (\ref{eq:TDX}). Unlike the TDX distribution in forward kinematics for an orbital angular momentum transfer $L=0$, it is no longer a peak distribution centered at the value of $T_p^{RL}$~\cite{nadasen_1989}. 

In Figure~\ref{fig:fig-TDX}(a) the experimental TDX is compared to the result of the distorted-wave impulse approximation~(DWIA) calculations using microscopic Reduced Width Amplitude (RWA) of the ground state of $^{10}$Be obtained through the THSR and AMD models.
Explicit formulae of ($p$,$p\alpha$) TDX calculated within the DWIA framework can be found in Eqs.~(4)--(7) of Ref.~\cite{Yoshida2022_Po}. See Ref.~\cite{Chant1977,Chant1983,Wakasa2017} for detailed DWIA description. 
TDX calculations have been performed for an incident energy of 150 MeV/nucleon used in the experiment. 
The proton optical potentials were deduced from the democratic parameterization of Dirac phenomenology~\cite{Cooper2009}. 
The global optical potential of Ref.~\cite{Avrigeanu94} is applied to the emitted $\alpha$.
As for the $p$-$\alpha$ elementary process, $p$-$\alpha$ differential cross section obtained by the folding model potential~\cite{Toyokawa2013} using the Melbourne $G$-matrix interaction~\cite{Amos2000} is adopted. 
For the structure calculation of the $^{10}$Be ground-state, the RWA used in the TDX calculation has been obtained from the approximation method described in~\cite{kanada-enyo_approximation_2014} for both THSR and AMD cases. The $^{10}$Be ground-state calculated within the THSR model was presented in Ref.~\cite{lyu2018}. It corresponds to a molecular configuration of the 2$\alpha$ core with two valence neutrons occupying ring-shape $\pi$ orbitals. The intrinsic proton and valence density distribution are shown in Fig.~\ref{fig:fig-TDX}(c) and (d), respectively. 
The $^{10}$Be ground-state from AMD model~\cite{kanada-enyo_structure_1999} exhibits very similar features. 

The shape of the experimental TDX distribution of $^{10}$Be$(p,p\alpha)^{6}$He(g.s.) is very well reproduced by both calculations as can be seen in Figure~\ref{fig:fig-TDX}. Furthermore, the normalization of the calculated distributions to the experimental one by a fitting procedure leads to normalization factors of 1.04(7) and 0.90(6) for the THSR and AMD, respectively, very close to unity. Given the strong dependence of the TDX on the $\alpha$-cluster wave-function reported in Ref.~\cite{lyu2018}, one can conclude that the present microscopic descriptions of the $^{10}$Be$(p,p\alpha)^{6}$He(g.s.) reaction allow an accurate reproduction of the data.

\begin{table}[htpb]
		\centering
            \setlength{\tabcolsep}{1.6mm}
		\caption{Comparison of the experimental and theoretical double-differential cross-sections for the ground state and 2$_{1}^{+}$ excited state transitions at quasifree conditions. Both $\sigma_{exp}$ and $\sigma_{th}$ are integrated over the angle bin size of $\pm 1 ^{\circ}$ for the quasifree angle pairs at $\theta_p=65^{\circ}$.}
		\label{tab:intCX65}
		\begin{tabular}{c c c c c}
		\hline
		\hline
		Final& $[\theta_{p}/\theta_{\alpha}]$ & $\text{DDX}_{exp}$& $\text{DDX}_{\text{THSR}}$ & $\text{DDX}_{\text{AMD}}$ \\
		  state & (deg) & (mb/$\text{sr}^2$) & (mb/$\text{sr}^2$) & (mb/$\text{sr}^2$) \\
		\hline
  		$^{6}$He(g.s.) &  $65^{\circ}/7.7^{\circ}$ & 23.6(28)& 22.7 & 25.9\\
		$^{6}$He(2$_{1}^{+}$)& $65^{\circ}/7.5^{\circ}$ & 17.6(30) & 5.2 &7.9\\
	
		\hline
		\hline
	
		\end{tabular}
\end{table}
The population of excited states in the $^{6}$He residues through the ($p$,$p\alpha$) reaction measures the contribution of $^6$He core-excited states in the ground state of $^{10}$Be. Extracting the corresponding cross-sections in addition to the one for the ground-state transition provides a stringent test to the structure models of $^{10}$Be.
Table \ref{tab:intCX65} gives the extracted double differential cross-sections (DDX) for the ground-state and 2$_1^{+}$ excited-state transitions at the quasifree condition obtained by integrating the TDX over the proton kinetic energy $T_p$. 
Due to the slight variation of the residue mass, there is a corresponding change in the quasifree angle pairs, while $\theta_{p} = 65^{\circ}$ remains unchanged for both cases.
The number of counts for the g.s. transition is extracted straightforwardly as only the g.s. of $^{6}$He is bound,
while that for the 2$_{1}^{+}$ excited state transition is obtained through a decomposition of the excitation energy spectrum, as shown in Fig.~\ref{fig:fig-Ex}(b).
The spectrum was fitted by two resonances modeled as the convolution of a Breit–Wigner distribution with a Gaussian function, taking into account the experimental resolution.
The amplitudes and energies ($E$) of the two resonances are treated as free parameters, while keeping the width of the 2$_{1}^{+}$ state (0.113 keV) fixed.
Two approaches for handling the width of the additional resonance were explored: energy-dependent $\mathit{\Gamma} (E)$ from the R-matrix method~\cite{Wuosmaa2017PRC,AKSYUTINA2009PLB} and the approach described in Ref~\cite{Matta2015PRC}(shown in Fig. 2(b)), yielding consistent results. 
The present spectrum is well reproduced by a fit with a resonance identified to the well-known 2$_{1}^{+}$ state and an additional resonance at 5.0(2) MeV, $\mathit{\Gamma}$~=~2.3(5)~MeV, consistent with Ref~\cite{mougeot_new_2012,Chernyshev2018KnE}.

The experimental and calculated DDXs are compared in Table \ref{tab:intCX65}. A very good agreement is obtained for the ground-state transition as expected from the TDX comparison in Fig.~\ref{fig:fig-TDX}(a). The DDXs of the transition to the 2$_{1}^{+}$ state are smaller than for the ground-state transition, in particular for the theoretical predictions.
The discrepancy between the experimental and theoretical values may arise from the inadequate treatment of time-dependent resonances in the structure models and the reaction models. The present THSR and AMD calculations described the 2$_{1}^{+}$ resonant state with boundary conditions as those of the bound states, resulting in lower predictions for cross-sections due to the different matrix elements between initial and final states.

\paragraph{Discussion}

The $\alpha$-cluster molecular structure of the $^{10}$Be ground-state within the THSR-based framework shown above (2$\alpha$ cores with the two valence neutrons occupying ring-shape $\pi$ orbitals) allows the reproduction of our experimental cross-sections very well. 
This configuration is spatially extended, although to a lower extent than for other beryllium isotopes because of the attractive effect of the $\pi$ neutrons compared to e.g. neutrons in $\sigma$ orbits. Consistently, the corresponding root-mean-square charge radius of $^{10}$Be is 2.31 fm, very close to the experimental value.  
To estimate the uncertainty from potential choices on cross-sections, we performed further TDX calculations using the Koning-Delaroche (KD) potential for $p-^{10}$Be and $p-^{6}$He\cite{Egashira2014PRC} and the $\alpha-^6$He optical potential was obtained by the nucleon-nucleus folding calculation using the KD potential and the phenomenological alpha density\cite{Koning2003NPA}. Both sets of potentials reasonably reproduced the experimental TDX, with the later calculation showing scaling factors of approximately 80$\%$ compared to the original calculation.

The sensitivity of the TDX of the $^{10}$Be$(p,p\alpha)^{6}$He(g.s.) reaction to the spatial extension of the $\alpha$ cluster wave-function, quantified by the intercluster distance has been clearly demonstrated in~\cite{lyu2018}, and the present results thus provide a direct evidence of the above molecular structure. 
To confirm that these conclusions are valid in our present experimental conditions (incident energy, finite angular acceptances), we performed the TDX calculation using the two "extreme" cases of a compact shell-model-like and a loosely bound gas-like configuration of $^{10}$Be nucleus described in Ref.~\cite{lyu2018}. 
Although unphysical, these states test the impact of $\alpha$-clusters with different spatial distribution on the TDX magnitude.
As can be seen in Fig.~\ref{fig:fig-TDX}(a), very large normalization factors are needed to match the magnitude of the data, far beyond the cross-section calculation uncertainty. 
It can be also noted that the TDX calculation corresponding to the gas-like configuration has a different shape compared to the experimental data. Altogether, our measurement validates the microscopic physical THSR $\alpha$-cluster wave function mentioned above.

The AMD approach has been successfully used to describe and establish low-lying molecular structures in light nuclei.
This framework provides a microscopic description of both single-nucleon properties as well as cluster structure, without assuming the cluster \textit{a priori}.
Within this framework, the $^{10}$Be ground-state is found to have a structure similar to the one extracted from the THSR model wave function in the region of interest. Namely, the RWA has similar behaviour at the surface region which contributes to the cross-section. Consistently, we find that the calculated TDX using AMD RWA shows an agreement of similar quality with the data as when using the THSR approach. 

\paragraph{Conclusion}
The alpha-cluster structure of an unstable neutron-rich nucleus, $^{10}$Be,
has been investigated by measuring for the first time the TDX of the ($p,p\alpha)$ reaction in inverse kinematics with a setup allowing inclusion of the recoilless condition. 
Double differential cross-sections to the ground and 2$_{1}^{+}$ states of the $^{6}$He residue have been extracted independently. Obtained data have been compared with cross-section calculations performed within a microscopic DWIA framework involving up-to-date $\alpha$-cluster wave functions, 
describing the ground-state in terms of a dumbbell shape 2$\alpha$ core (with moderate extension) surrounded by the two valence neutrons occupying the $\pi$ orbit.
There is a remarkable agreement in both shape and magnitude between the experimental and calculated TDX for the $^{10}$Be($p,p\alpha)^{6}$He(g.s.). 
Due to the previously established sensitivity of the TDX to the extension of the alpha wave-function in the ground-state of $^{10}$Be, 
our results provide direct experimental evidence of the above molecular structure of $^{10}$Be implemented in the THSR approach, and validated by the general AMD framework. 
Concerning the double differential cross-sections to the g.s. and 2$_{1}^{+}$ states of the residue, a good agreement between calculations and experimental results is found. 
A consistent picture is then obtained. In the near future, based on the present work further studies will lead to the understanding of the evolution of the alpha cluster structure in the ground-state of neutron-rich nuclei with increasing number of valence neutrons. Besides, it is also planned to apply knockout reactions in inverse kinematics to investigate the formation of other types of clusters in the ground-state of nuclei away from the stability valley.

\begin{acknowledgments}
\paragraph{Acknowledgements} We would like to express our gratitude to the RIKEN Nishina Center accelerator staff for providing the high-intensity primary beam and to the BigRIPS team for their efforts in preparing the secondary beams.
P.\,J.\,L. wish to acknowledge the fruitful discussions with Z.\,H. Li, Z.\,Y. Tian, and Q. Zhao.
J.\,L. acknowledges the support from the Research Grants Council of the Hong Kong Special Administrative Region, China (RGC/GRF, HKU 17304918).
X.\,X.\,X. acknowledges the support from the Strategic Priority Research Program of Chinese Academy of Sciences under Grant No. XDB34010300.
K.\,O. acknowledges support from the Grants-in-Aid of the Japan Society for the Promotion of Science (Grants No. JP20K14475 and No. JP21H00125). K.\,O. and T.\,U. acknowledge support from the Grants-in-Aid of the Japan Society for the Promotion of Science (Grant No. JP21H04975).
L.\,S. acknowledges the support from the Institute for Basic Science (IBS-R031-D1).
T.\,N. and K.\,Y. acknowledge the support from the JSPS KAKENHI(Grant Nos. JP18H05404 and JP21H04465). Z.\,H.\,Y. acknowledges the support from the National Key R$\&$D Program of China (Grant Nos. 2022YFA1605100, 2023YFE0101500) and the National Natural Science Foundation of China (Grant No. 12275006). D.\,B. acknowledges the support from the Invitation Fellowship Program for Research in Japan of the Japan Society for the Promotion of Science (Grant No. L11707). P.\,J.\,L. acknowledges the support from China Postdoctoral Science Foundation(Grant No. YJ20210186).
M.\,L. acknowledges the supported from National Natural Science Foundation of China (Grants No. 12105141), and Jiangsu Provincial Natural Science Foundation (Grants No. BK20210277). Y.\,T. acknowledges the support from JSPS Grant-in-Aid for Scientific Research Grants No. JP21H01114.
\end{acknowledgments}

\bibliographystyle{apsrev4-2.bst}
\bibliography{example.bib}

\end{document}